# Fog Computing Approaches in Smart Cities: A State-of-the-Art Review

Maryam Songhorabadi, Morteza Rahimi, Amir Mahdi Moghaddam Farid, and Mostafa Haghi Kashani*

**Abstract**— These days, the development of smart cities, specifically in location-aware, latency-sensitive, and security-crucial applications (such as emergency fire events, patient health monitoring, or real-time manufacturing) heavily depends on a more advance computing paradigms that can address these requirements. In this regard, fog computing, a robust cloud computing complement, plays a preponderant role by virtue of locating closer to the end-devices. Nonetheless, utilized approaches in smart cities are frequently cloud-based, which causes not only the security and time-sensitive services to suffer but also its flexibility and reliability to be restricted. So as to obviate the limitations of cloud and other related computing paradigms such as edge computing, this paper proposes a systematic literature review (SLR) for the state-of-the-art fog-based approaches in smart cities. Furthermore, according to the content of the reviewed researches, a taxonomy is proposed, falls into three classes, including service-based, resource-based, and application-based. This SLR also investigates the evaluation factors, used tools, evaluation methods, merits, and demerits of each class. Types of proposed algorithms in each class are mentioned as well. Above all else, by taking various perspectives into account, comprehensive and distinctive open issues and challenges are provided via classifying future trends and issues into practical sub-classes.

**Index Terms**— Fog Computing, Internet of Things (IoT), Systematic Review, Smart Cities, Urban.

———————————— ◆ ————————————

## 1 INTRODUCTION

EXTENDING from the cloud, fog computing, a geographically distributed paradigm, deploys computing and networking power into the network edge, nearer to both end-users and IoT devices, thanks to the support of wide-spread fog nodes [1]. In cloud-only architectures, most data that require processing, analysis, and storage are sent to the cloud data servers that might adversely affect latency, security, mobility, and reliability. By the emergence of location-aware and delay-sensitive applications (esp. in smart cities and smart buildings), the cloud alone faces severe issues to fulfill the extremely-low latency needs of these applications. However, the close vicinity of the fog layer to the Internet of Things (IoT) devices of a smart city can significantly reduce latency and satisfy the requirements of extremely-low latency applications [2]. Indeed, fog computing is a supportive layer for the cloud and is always in interaction with it, which creates a new generation of applications and services. Fog computing can be utilized in quite a few related domains, including healthcare [3], big data analytics [4], IoT [1], smart homes [5], smart cities [6], wearable/mobile computing [7].

The main goal of the smart city is to enhance the quality and convenience of urban life, which is an integral part of politics in contemporary big cities. With the growth of the urban population and extreme penchant for improving quality of urbanization, smart cities emerged, aiming at providing comfortable services that can be sustainable, available, rapid, and secure. In a smart city, transportation services such as intelligent vehicle commutation [8], along with energy efficiency [9], health-care services [10], network types [11], and citizens telecommunications [12] are the domains that can take advantage of fog computing and its benefits.

Nowadays, quite a few services and applications, namely self-driving vehicles, video surveillance, online data analytics, live streaming, security monitoring, health emergencies, and real-time manufacturing, are offered in smart cities, among which the security-crucial, latency-sensitive and location-sensitive services are of notable consequence [2, 13]. Any delay in receiving, sending, or processing these data may lead to irrecoverable damages. Besides, fog computing can considerably decrease the communication/computation latency and increase security through wide-spread fog nodes with much closer physical distances. Hence, to face the indispensable requirements of smart cities and support cloud computing, fog technology is an optimal solution, specifically to deal with the current smart city, IoT complexities [14].

To date, only one systematic review has assessed fog-based applications in smart cities. With respect to the authors, their work suffers from several limitations that should be considered. *First and foremost, the authors have considered fog computing and edge computing precisely the same in their search string, yet there are several differences between them* (explained in Section 2.1.1). *Secondly, the authors overlooked some synonyms in their search string (e.g., urban) by*

---

- *Maryam Songhorabadi and Amir Mahdi Moghaddam Farid are with the Department of Computer Engineering, South Tehran Branch, Islamic Azad University, Tehran, Iran. E-mail: st_m_songhorabadi@azad.ac.ir, st_am_farid@azad.ac.ir.*
- *Morteza Rahimi is with Young Researchers and Elite Club, Shahr-e-Qods Branch, Islamic Azad University, Tehran, Iran. E-mail: m.rahimi@qodsiau.ac.ir*
- *Mostafa Haghi Kashani is with the Department of Computer Engineering, Shahr-e-Qods Branch, Islamic Azad University, Tehran, Iran. E-mail: mh.kashani@qodsiau.ac.ir.*
- *\*The corresponding author is Mostafa Haghi Kashani.*
- 

*which several papers (about fog-based urban sound classification, urban transition, urban traffics management, urban IoT data streams and so forth) were missed.* Motivated by these limitations and in order to resolve them, we arrange a distinctive SLR about fog-based smart cities *with a notable emphasis on fog computing rather than fog/edge computing.* An SLR recognizes, distinguishes, categorizes, and incorporates a comparative review of researches and allows knowledge to move into the community of researches [15].

More precisely, in order to provide a methodological overview of current studies, we intend to answer several questions that are elaborated upon hereunder:

- What are the primary crucial motivations behind applying fog computing approaches in smart cities?
- What are the existing techniques, approaches, and tasks, which enable fog computing approaches to be utilized in smart cities?
- What are the existing tools, evaluation factors, merits, and demerits of studies?
- What evaluation methods are applied in fog-based smart cities?
- What are the prevalent algorithm types applied in fog-based smart cities?
- What are the prevailing themes of research studies? More importantly, what pivotal factors should make subsequent research trends in fog-based smart cities?

Via following guidelines in [16, 17], we aim to recognize systematically and categorize taxonomically state-of-the-art researches available on fog-based smart cities along with providing an abreast comparison to investigate restrictions and potential of existing studies. This strategy supplies a systematic literature review of existing research, concentrating on presented approaches, methods, and solutions in fog-based smart cities. Accordingly, 32 studies are chosen, categorized, and compared, applying inclusion/exclusion criteria and selection stages (see Section 4). These selection stages and methodologies are purified and derived from a qualitative evaluation of the mentioned researches, some renowned references [18, 19], and our previous experiences in conducting SLRs [5, 20].

Furthermore, via providing an exhaustive search string, this paper investigates state-of-the-art fog-based approaches in smart cities more comprehensively. Our systematic review in this scope can extensively help researchers to apply innovation and advances in smart cities that *specifically take advantage of fog computing*. This work also categorizes the available approaches in fog computing based smart cities to perform a better assessment and shed light on the path of future researches. In particular, the outcomes of this SLR are advantageous for:

- Studies in fog and cloud computing, who requires an identification of relevant and related researches.
- Researches engrossed in grasping the available approaches and methods of fog computing in smart cities as well as their tools, evaluation factors, and future directions.

The rest of the papers is arranged as follows. The background is provided in Section 2. Section 3 includes related work and motivation. In Section 4, the research methodology has covered our paper selection procedure. The classification of the selected papers is represented in Section 5. Section 6 provides an analysis of the results. Section 7 describes the open issues and future works. Section 8 explains the threats to the validity of the review. Ultimately, Section 9 discusses the conclusion.

## 2 BACKGROUND

The overview of fog computing in smart cities and its significance in bringing facilities to cities is offered in this section. In the following, the concepts and descriptions have presented:

### 2.1 Fog Computing

The very first concept of fog computing was introduced by the computer information system company (Cisco) in 2012 [1]. According to OpenFog Consortium [21], fog computing is "a horizontal, system-level architecture that distributes computing, storage, control, and networking functions closer to the user along a cloud-to-thing continuum." The word "horizontal" refers to the platform that lets computing processes and functions be distributed between various platforms or even industries. Furthermore, fog computing supplies a platform that is flexible enough to meet the needs of users and operators [2]. Several exclusive, beneficial attributes of fog are as follows [22]:

**Security:** In the light of transmitting data and information in a remarkable short distance along with utilizing the same policy, procedure, and control for fog nodes, security vulnerabilities and risks decrease, and security monitoring and authentication methods perform as efficiently as possible.

**Latency:** Fog environment with widely distributed fog nodes, tremendously reduce the latency by virtue of closeness to end-devices. Hence, fog computing is axiomatically efficient in time-sensitive and also location-sensitive applications.

**Efficiency:** The nearer vicinity to end-devices of the endpoints enable fog computing to be more nearly integrated with end-device systems to improve overall performance and efficiency of the system. It is incontrovertibly a prominent characteristic for performance-crucial and cyber-physical systems.

**Cognition:** Vast spread fog nodes of fog architecture, can best recognize where to perform the storage, control functions, and computing by virtue of awareness of user requirements originated from being geographically distributed.

More importantly, three cloud computing major service models, including SaaS (software as a service), PaaS (platform as a service), and IaaS (infrastructure as a service), can also be implemented in the fog environment [14]. The differences between fog and other computing paradigms are elaborated upon hereunder [23]:

- Delay-sensitive and location-awareness: In a system where a fog node communicates with other fog nodes, fog computing efficiently suggests the lowest delay possible for path and response or analyze the infor-mation way much faster than centralized cloud computing. It is because fog nodes are aware of their logical location in the whole network.
- Real-time response: Fog environment carry out data analytics at the edge of the network by which time- and latency-sensitive control functions and applications occur in a dramatically closer distance.
- Heterogeneity: Different network types have different data processing or collections, and fog computing supports them all.
- Federation and interoperability: Supporting multiple services need the collaboration of varied providers. Therefore, services should be federated all over the amplitude.
- Geographical distribution: In contrast to the centralized structure of the cloud, fog computing is geographically distributed.
- Fog nodes agility and scalability in clusters: Fog computing at clusters and their levels, supports resource integration, elastic computing, network condition changings, and data-load variations that show some of the adaptive operations.

*2.1.1 Fog vs. Edge*

Fog computing and edge computing are very often mistakenly confused with each other, yet there are prominent differences between them. Running applications in a multitier architecture, fog computing meshes, and decouples the software and hardware functions through which dynamic reconfiguration is possible for various applications even when they are performing transmission services or intelligent computing. Edge computing, on the other hand, via running applications in a fastened location, supplies transmission service in a direct manner. Indeed, fog environment is highly hierarchical, while edge computing frequently tends to be restricted to a handful number of endpoint appliances and devices. Furthermore, besides networking and computation, the fog layer addresses data-processing, control, and storage acceleration [23].

*2.1.2 Fog Architecture*

Fog computing is a horizontal multitier architecture that expanded in the network topology with distributed storage, networking, and control-functions [14, 21]. As shown in Fig. 1, fog computing has a three-layered architecture. We have described every layer below:
- **Cloud layer:** In this layer, quite a few centralized storing and computing centers are gathered which analyze, store, and process data with high performance.
- **Fog layer:** In this layer, a large number of fog nodes are located, near the lowest layer, the user device layer. Processing data with the aim of increasing locality is the fog layer's responsibility. Physical devices such as gateways, routers, virtual machines, and switches are broadly distributed between the end-device layer and cloud layer.
- **User device layer:** This layer contains various devices that bring comfort to people's lives. Many cell phones, vehicles, sensors, cameras, etc., exist in this layer to transfer sensed data to the upper layers to be processed and stored.

In this three-tiered architecture, every user device or sensor is connected to fog nodes via wireless connections such as Bluetooth, 5G ($5^{th}$ generation), 4G ($4^{th}$ generation), Wi-Fi, ZigBee, or wired technologies. Moreover, fog nodes have the potential to make multiple connections with each other by wired or wireless technologies [2].

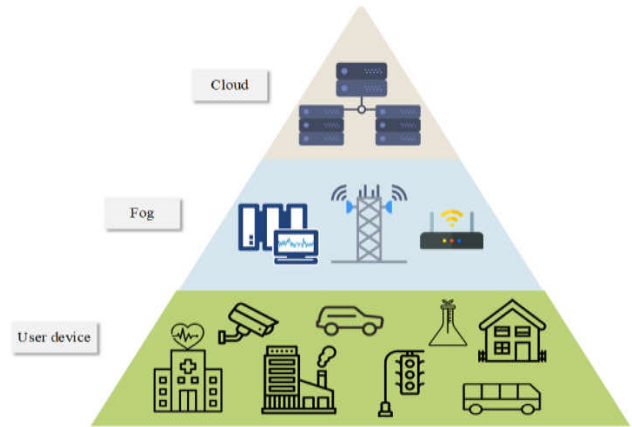

Fig. 1. Fog-based smart cities three-layered architecture.

*2.1.3 Fog Nodes*

In sharp contrast to centralized network architectures that network load adversely rises, through fog networking, the collected information is sent to highly distributed fog nodes in which computation, analysis, and storage is performed. In fog structure, there are virtual or physical components such as virtual machines, switches, servers, gateways, or routers, which are parts of the core layer [21]. They are responsible for providing a sort of information management and communication services or applications between two layers, user-device and cloud layer. Fog nodes present the following attributes and characteristics in a fog environment [23, 24].
- Heterogeneity: The ability to support and process different forms of data generated by various types of networks.
- Being manageable: The ability to run regular and routine functions and operations automatically due to being orchestrated and managed by sophisticated systems.
- Autonomy: Independent operations and local decisions can be performed by fog nodes at every level of nodes or clusters.
- Being programmable: The ability to be programmed and developed by domain experts, network developers, users, and providers.
- Hierarchical structure: Fog nodes, along with providing various service operations, support hierarchical clustering while started to work together.

**2.2 Smart Cities Genesis**

The vast areas we known as cities are the living areas for the majority of the worlds' population. In the early 20th century, imagining a city handling intelligence technology was something like a dream, but the computing power and networking grew over time, and humans applied it wisely enough to prove that a smart city is no longer science fiction or a dream. The intelligent communication and information technologies got converged and brought an intelligent urban life, which is entirely different from everything we have experienced until now. Smart cities not only help conventional functions rendering traffic systems, houses and buildings, and people but also enable analyzing, monitoring, and planning to make a better quality of life for citizens [30]. Recent smart city definitions, define smart cities as urban agglomerations in which a broad spectrum of IoT devices and sensors are used to collect data and apply them to handle resources effectively [32]. However, cases such as traffic management, real-time manufacturing, self-driving vehicles, fire emergencies, and health monitoring of patients are the issues that cannot be solved by smart city equipment alone. Hence, applying security-crucial, location- and time-sensitive approaches of fog computing in the smart city brought favorable outcomes to cities.

TABLE 1
RELATED RESEARCHES IN FOG-BASED SMART CITIES

| Review type | Study | Publication year | Main focus | Methodology | Taxonomy | Open issue | Time range |
|---|---|---|---|---|---|---|---|
| Survey | [25] | 2017 | Fog-based smart cities | Not clear | No | Presented | Not mentioned |
|  | [26] | 2020 | Fog-enabled sustainable smart cities | Not clear | Yes | Presented | Not mentioned |
|  | [27] | 2019 | Fog-based 5G-enabled smart cities | Not clear | No | Presented | Not mentioned |
|  | [28] | 2020 | Edge-enabled smart cities | Not clear | Yes | Presented | Not mentioned |
|  | [29] | 2017 | Cloud-based smart cities | Not clear | No | presented | Not mentioned |
|  | [30] | 2019 | Smart city challenges and technologies | Not clear | Yes | Not presented | Not mentioned |
| SLR | [31] | 2020 | Fog applications in smart cities | Clear | Yes | Presented | 2013-2018 |
|  | **Our paper** | - | Fog-based approaches in smart cities | Clear | Yes | Presented | 2013-2020 |

In a fog-based city supporting IoT applications, multiple services can be enabled to enhance the quality of life. Given the challenges and issues that smart cities face, the researches can be dedicated to exploiting from hybrid cloud/fog computing, IoT, and mobile devices technologies. In particular, transportation services such as intelligent vehicle commutation [8], along with energy efficiency [9], health-care services [10], network types [11], and citizens telecommunications [12] are the domains that can take advantage of fog computing and its benefits.

## 2.3 Metrics Definition

Different metrics need to be compared to evaluate different procedures in the fog based smart cities. These comparisons can help identify these procedures and find their strengths and weaknesses. This paper compares the most prominent metrics of papers reviewed by fog-based smart cities. These metrics are listed below:

- Scalability: The ability to extend fog-based technologies in services, nodes, and applications
- Response time: The interval between sending a request and receiving a response from a fog node
- Security: The activities that prevent damage to data or services and retain the usability of information or technology (hardware, software).
- Energy: The amount of utilized power in smart cities network.
- Cost: What a component or supplicant pays in return for a need.
- Throughput: Is the quantity that shows the amount of processed information in a specific time in a system.
- Latency: the transmit time that a packet needs to spend.

## 3 RELATED WORK AND MOTIVATION

The relevant papers of fog-, edge-, and cloud-based smart cities are gathered together in this section. They have been reviewed and discussed in the following paragraphs. According to a systematic search (Section 3.2), we investigate available reviews and surveys in related domains (Section 3.1).

### 3.1 Relevant reviews on Fog and Smart Cities

During recent years, several reviews and surveys have concentrated on edge-based and cloud-based smart cities. Additionally, a few reviews have investigated fog computing applications in smart cities, among which only one review has followed systematic structure (mentioned in Table 1). As previously mentioned in Section 1, in order to resolve the limitation of that SLR and concentrate on delay- and location-sensitive applications of smart cities through fog computing, we determined to investigate existing related surveys to arrange this SLR. These studies summarize in Table 1, as secondary studies, based on survey and SLR.

#### 3.1.1 Surveys

Perera, et al. [25] investigated techniques to solve the issues of fog computing-based smart cities. In particular, several scenarios have been described, and several key features have been identified to analyze the various types of fog computing to recognize the needs of a platform. The high impact of fog platforms for IoT in order to create sustainable infrastructure in smart cities has been clearly demonstrated. The authors have identified key features that fog platforms should have and several open issues in their implementation. Nonetheless, this paper did not follow a systematic structure, and the paper choosing process is not determined.

Zahmatkesh and Al-Turjman [26] investigated fog computing applications in smart cities and discussed caching techniques in the IoT communications of smart cities. Moreover, the paper reviewed ML/AI (machine learning/artificial intelligence) methods and UAVs (Unmanned Aerial Vehicles) in catching data for fog-enabled IoT smart systems. Although the authors investigated future trends and challenges, the lack of systematic structure causes their paper selection process left unspecified. Also, Khan [27] overviewed fog computing approached in 5G-enabled smart cities. Although the authors created a connection between three essential technologies, including 5G, smart cities, and fog/edge computing through a framework, open issues and research methodology, were left unspecified. Similarly, Khan, et al. [28] surveyed edge-enabled smart cities in which they highlighted the evolution of edge environment paradigm. Although the authors proposed a comprehensive survey with clear taxonomy, the year of reviewed papers was left unspecified due to a lack of systematic structure.

Petrolo, et al. [29] browsed the annotation of the cloud sensors, and also investigated advanced services that can be considered and implemented by IoT and CoT (cloud of things) cohesion. This survey provided extremely concise open issues and challenges and did not present a taxonomy or classification. Furthermore, the lack of systematic structure causes an undefined paper selection process.

Sánchez-Corcuera, et al. [30] represented new issues and challenges of the smart cities. In this paper, some methodologies and technologies have been reviewed. The domain classification that is specified in this review is well-descried. However, for implementing smart cities, many factors are still neglected, like providing primary re-requirements and services and furthering the progress of environmental protection and sustainability along with a proper roadmap that can be a guide in the field of smart cities. Also, the review approach was not systematic; the process of selecting papers is not indicated.

### 3.1.2 Systematic Review

Javadzadeh and Rahmani [31] presented an SLR on fog-based applications to optimize resource management and provide real-time services in smart cities. Although this review peruses a systematic structure, the open issues section is arranged briefly and limited, and the recently published papers are neglected. *Moreover, the word "urban" was overlooked in their search string by which several papers were missed. Nonetheless, we considered a more exhaustive search string with a notable emphasis on fog computing.*

### 3.1.3 Concluding Remark

As for the research methodology and search string, the work of Javadzadeh and Rahmani [31] is the closest to ours; nonetheless, we provide a more comprehensive search string (including word "urban") and put notable emphasis on fog computing rather than fog/edge computing due to their prominent distinctions (see Section 2.1.1). Except for one of the works, the time range of the papers was not specified, which covered the studies up to 2018. However, we review recently published papers. Furthermore, some papers did not propose any taxonomy, while we propose a practical taxonomy in our paper. The research methodology was left unspecified in almost all of the studies; hence, the paper selection procedure is undetermined. In sharp contrast, we propose an SLR in which research methodology is well-determined. Last but not least, most of the papers had not perused evaluation methods and used tools in smart cities. Nevertheless, in this SLR, we compare evaluation methods and applied tools along with their algorithm types.

## 3.2 The Motivations Behind Another Review

The requirement for an SLR involves recognizing, categorizing, and comparing available studies in fog-based smart cities. So as to find the relevant reviews, we searched the Google Scholar, IEEE Xplore, ScienceDirect, ACM, and Springer (on 5/28/2020) with the following search strings:

*("fog" <OR> "edge") [AND]*

*("city" <OR> "cities" <OR> "urban") [AND]*

*("Review" <OR> "SLR" <OR> "Systematic Mapping" <OR> "Survey" <OR> "Secondary Study" <OR> "Overview" <OR> "Challenges" OR "Trends" <OR> "State of The Art")*

In Table 1, the main concentrations and information about each related paper are demonstrated. Given the importance of applying fog computing approaches in a smart environment, specifically in delay- and location-sensitive applications, integration of current evidence to smart cities is opportune and timely.

## 4 RESEARCH METHODOLOGY

A systematic literature review [16] diminishes bias and pursues a rigorous and valuable sequence of several methodological stages to literature, in sharp contrast to a non-structured review process [17]. *SLR methods rely on clear-cut and well-defined researches and derived information of practical results* [18], *as depicted in* Fig 2. We pursued a three-stage review procedure consisting of the *planning* stage, *conducting* stage, and *documenting* stage according to guidelines in [16].

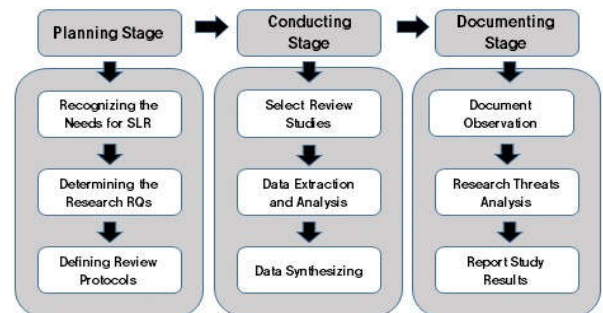

Fig. 2. Overview of the research methodology.

### 4.1 Planning the Systematic Review

The planning stage commences with recognizing the needs and motivations for a systematic review and ends in defining review protocols.

*1) Recognizing the needs and motivations:* The motivations and needs, along with the contribution of this Systematic review is justified and recognized in Section 3. Furthermore, the research questions (RQs) are specified in Table 2, which greatly help in evaluating and defining the review protocol for conducting the SLR.

*2) Determining the research questions:* According to our motivation and limitations of related works, we define the research questions. The answers give us a science-based review of fog computing approaches in smart cities. Our six research questions, along with their motivations, determined in Table 2.

*3) Defining review protocols:* According to objectives, we determine the review scope to create the search string for available literature extraction methodically. Also, a protocol is developed for SLR by pursuing [16] and our previous experience with systematic literature reviews [5, 20, 33]. As recommended by Brereton, et al. [16], we assessed the protocol before execution. We have asked two external SLR experts for feedback, who had broad experience in arranging SLRs in the domain of fog computing, cloud computing, and IoT, so as to assess the presented protocol. In this regard, we refined our proposed review protocol according to the external expert feedback. Additionally, we carried out a pilot study (approximately 20 percent) of our included researches. Carrying out a pilot study significantly decreases the bias between the researchers and enhance data collection scheme. We developed the scope of the review, enhance search strategies, and improve the inclusion/exclusion criteria during the pilot studies.

TABLE 2
RESEARCH QUESTIONS AND MOTIVATION BEHIND THEM

| Research Questions | Motivation |
| --- | --- |
| **RQ1:** What are the primary crucial motivations behind applying fog computing approaches in smart cities? | The aim is to elucidate security-crucial, location- and time-sensitive applications of fog computing in smart cities. |
| **RQ2:** What are the existing techniques, approaches, and tasks, which enable fog computing approaches to be utilized in smart cities? | The aim is to determine and compare available approaches and techniques of fog in smart cities in a systematic manner. |
| **RQ3:** What are the tools, evaluation factors, merits, and demerits of studies? | The aim is to recognize and compare applied metrics, tools, pros, and cons of existing fog-based smart city studies. |
| **RQ4:** What evaluation methods are applied in fog-based smart cities? | The aim is to specify the most prevalent evaluation methods in fog-based smart cities. |
| **RQ5:** What are the prevalent algorithm types applied in fog-based smart cities? | The aim is to find the best optimization methods and algorithms in smart cities that benefit from fog computing. |
| **RQ6:** What are the prevailing themes of research studies? More importantly, what pivotal factors should make subsequent research trends in fog-based smart cities? | The aim is to grasp and define the research gaps that need to be identified and potential future trends in this field. |

## 4.2 Conducting the systematic review

The conducting stage consists of study selection, outcomes in extracted data, and information synthesizing.

### 4.2.1 Study Selection

We specified search terms using [16] and guided by our proposed motivations and research questions in Section 3.1.3 and 4.1. The paper selection process in this stage is accomplished in 2 steps, containing I) initial selection and II) final selection. The first stage of the research process was searching through well-known publishers such as IEEE, ACM, Springer, ScienceDirect, Wiley, and Google scholar. The following string which is derived from keyword searching and their synonyms searched in the aforementioned eminent databases:

"fog" [**AND**] ("city" <**OR**> "cities" <**OR**> "urban")

As mentioned in Section 3.1.3, *this string is more comprehensive compared with the closest work to ours, by virtue of covering papers about fog-based **urban** sound classification, **urban** transition, **urban** traffics management, **urban** IoT data streams and so forth*. Most importantly, we arrange a distinctive search string *with a notable emphasis on fog computing rather than fog/edge computing due to their prominent differences* (see Section 2.1.1).

• *Initial selection:* According to keyword searching, 726 papers extracted from 2013 to 2020 (the year 2013 was selected because no earlier study was found related to our topic). Then via applying inclusion/exclusion criteria (illustrated in Table 3), 98 papers have selected. The concentration of our search process was on the papers published by IEEE, Springer, ScienceDirect, Wiley, and ACM. For almost 25% of the researches, we could not make any decision. In these cases, we were carrying on to final selection associated with thoroughly investigating the full text.

• *Final selection:* This stage is based on an authentication scan of the researches. After performing this stage, 32 studies were chosen for further investigation. Only journals and conferences are included as final studies to be reviewed. The exclusion list is shown in Table 3.

TABLE 3
RESEARCH CRITERIA

| | Research Criteria | Rationale |
| --- | --- | --- |
| Inclusion | • Researches in the form of a scientific and academic peer-reviewed paper. | These researches guarantee a favorable amount of quality and an ample amount of content. |
| | • Researches that lucidly present techniques, experiences, solutions, and assessment to facilitate fog-based approaches in smart cities. | We aim to investigate clearly defined metrics, solutions, and analyses in fog-based smart cities. |
| Exclusion | • Non-peer reviewed researches, non-English manuscripts, and white papers. | Quite a few technical reports and white papers are available; however, their quality is situational. |
| | • Book chapters, books, Thesis. | These studies are often published in conferences and journals. We already included relevant papers. |
| | • Editorials, short papers, or abstracts | Due to the lack of providing an ample amount of content and considerable solutions, we exclude them. |
| | • Researches do not lucidly present techniques, solutions, or assessments to facilitate fog-based approaches in smart cities. | These researches investigate related domains that do not describe fog-based approaches in smart cities. For instance, researches are related to edge computing. |

### 4.2.2 Data Extraction and Synthesis

We arranged a systematic format according to selection stages and strategies provided in guidelines by [16, 18]. Based on Table 4, we have shown the number of selected journals and conferences separately. 10 journals and 9 conferences are published by IEEE, which has the highest number of published papers, among others. ScienceDirect published 7 journals, and Springer has published 2 journals and 2 conferences equally. Ultimately, 1 journal is published in Wiley, and 1 conference paper is published in ACM.

TABLE 4
DISTRIBUTION OF PAPERS ON DIGITAL LIBRARIES

| Databases | Number of selected papers | Count |
| --- | --- | --- |
| IEEE | Journal | 10 |
| | Conference | 9 |
| ScienceDirect | Journal | 7 |
| Springer | Journal | 2 |
| | Conference | 2 |
| Wiley | Journal | 1 |
| ACM | Conference | 1 |

# 5 CLASSIFICATION OF THE SELECTED PAPERS

In this section, we have proposed a taxonomy of the literature based on the reviewed papers. Since the literature on fog computing and smart cities are hugely diverse, organizing and systematically reviewing researches is not easy. According to the proposed approaches of reviewed studies, we understood that all of them were related to the services, resources, and applications. Therefore, based on the extracted information, the 32 chosen papers, selected by stages in the previous section, are classified into three classes, consisting of services-based, resource-based, aplication-based. Although one of the primary purposes of this SLR is to present the most practical taxonomy, other classifications might be feasible. An illustrated outline of our proposed taxonomy, along with sub-classes in each class, is shown in Fig. 3. Moreover, the main specifications, differences, strengths, and weak points of each investigated paper are described in the remaining part of this section.

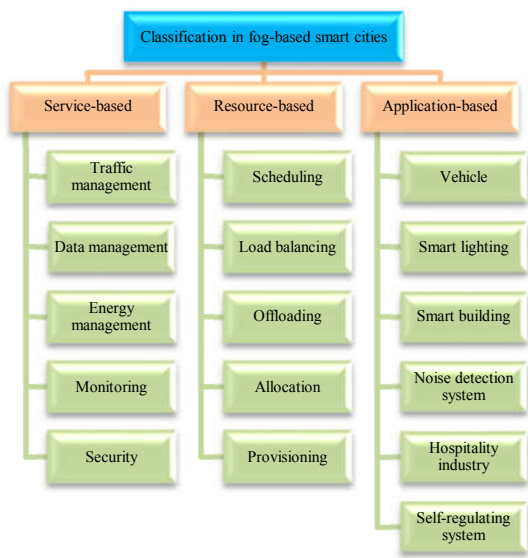

Fig. 3. Pictorial taxonomy of fog-based smart cities, including sub-classes.

## 5.1 Service-Based Approaches

This subsection introduces service-based approaches in fog-based smart cities. The selected papers in this category are developing or introducing services such as data management [34], [4], [35], [36], [37], [38], traffic management [39], [40], [41], energy management [42], [43], monitoring [44], [45], and security [46], which brings more convenience to smart cities. It will indicate the directions and trends of research on fog-based smart cities. Also, Table 5 displays a comparison of the main focus, evaluation method, algorithm type, tool, pros, and cons of each reviewed papers.

### 5.1.1 Review of the Service-Based Approaches

Tang, et al. [34] introduced a fog-based architecture in a smart city were helped to integrate a vast number of infrastructure services and analyze data. The architecture included layers, and each node at any layer quickly controls the security of essential infrastructure parts. However, the disadvantage of this paper was that it is not clear how available this architecture might be. Moreover, the reliability and security of the system were not thoroughly evaluated. Tang, et al. [4] introduced a distributed fog-based hierarchical architecture to support services and integration of a large number of components for analyzing a substantial amount of data in smart cities. One significant point in this structure was that all layers ran the latency-sensitive program and provided a mechanism to check critical components. A multitier fog-based model for analyzing a massive amount of data was introduced by He, et al. [35] that can alleviate probable issues of delay in responding and the infrastructure of computation. Although this model is scalable, but it was not mentioned how secure and reliable it could be.

Sinaeepourfard, et al. [36] presented F2c2C-DM (fog-to-cloudlet-to-cloud data management) architecture in fog-based smart cities. F2c2C-DM was designed with cloud and fog to organize enormous data generated by IoT sensors and devices in smart cities. The main advantage of this architecture is the benefit of distributed and centralized management. Badidi [37], proposed a fog-based architecture for urban environments that worked with data management in a city where different equipment and device generated. The architecture consisted of multiple parts to make a pipeline, and it let a considerable amount of data pass through IoT gateways to be processed. However, the security and reliability of this architecture were left uninvestigated. Eventually, on data acquisition for mobile crowdsensing, Fiandrino, et al. [38] presented a new framework for received data acquisition in mobile crowdsensing through a fog platform in smart cities. The assessment of performance was done for a high number of contributors, and it increased over 20% of the mediocrity of users. In contrast, the parameters such as reliability and security were not investigated, and the scalability was not fully covered.

Brennand, et al. [39], proposed an architecture for smart transportation system called FOREVER (FOg RoutE VEhiculaR) to mitigate the traffic agglomeration. The architecture was proposed to help to manage the traffic in networks of vehicles. FOREVER was a fog-based paradigm to modify and process the recommended ways. However, the mentioned mechanism had not specified how heavy the traffic was in order to notice vehicles of traffic agglomeration. On vehicular network routing, Lu, et al. [40] proposed IGR (improved geographic routing) based on position routing for the smart transportation. In order to use the benefits of fog computing, IGR used vehicular fog computing (VFC) for better communication and computation. Ning, et al. [41] introduced a three-layered VFC model for fog and vehicle integration in a smart city to manage traffic and alleviate the response time. The benefits of this model were: 1) decentralized network structure helped to handle the information processing, 2) fog and cloud integration for efficient offloading traffic on the network and reducing the burden on it, 3) fog and cloud nearness to terminal decreased response delay.

Butt, et al. [42] presented a three-layered architecture including fog, cloud, and end-devices in which several meta-heuristic algorithms namely, Genetic Algorithm (GA), and

TABLE 5
A COMPARISON OF SERVICE-BASED APPROACHES
In the evaluation method column, Prototype=> P, Simulation=>S, Real-testbed=>R, and Design=> D
In the algorithm type column, Non-Heuristic=>N, Metaheuristic=>M, and Heuristic=>H.

| Group | Ref | Main Focuses | Evaluation Methods | Algorithm Types | Tools | Advantages | Disadvantages |
|---|---|---|---|---|---|---|---|
| Data | [34] | Big data analysis using hierarchical fog computing architecture in smart cities | P | - | Hadoop | • High scalability<br>• Low response time | • Low availability<br>• Low security |
| | [4] | Big data analysis in fog-based smart cities | P | - | Not mentioned | • High scalability<br>• Low latency<br>• Low response time | • Low security<br>• Low availability |
| | [35] | Data analysis for large-scope smart cities based on multi-layer fog computing | S | H | GTgraph, Spark dataset generator | • High scalability<br>• High availability<br>• Low job completion time | • Low security<br>• Low reliability<br>• High complexity |
| | [36] | An architecture for data management in fog and cloud-based smart cities | D | - | Not mentioned | • High security<br>• High scalability<br>• Low latency<br>• Low network load | • Low reliability |
| | [37] | Real-time processing in fog-based urban data streams | P | - | Apache Kafka, Apache Storm, ThingsBoard | • Real-time responding | • Low scalability<br>• Low availability |
| | [38] | A data acquisition framework in mobile crowdsensing for fog-based smart cities | S | - | Google Transit Feed | • Low energy<br>• Low cost<br>• High sustainability | • Low availability<br>• Low extensibility |
| Traffic | [39] | A mechanism for traffic management based on fog computing | S | - | OMNeT++, SUMO, TraCI | • Energy-efficiency<br>• Low latency<br>• Low cost | • Low availability<br>• Low scalability |
| | [40] | Introducing a vehicular routing scheme for fog-based smart urban | S | H | NS-2 | • Low cost<br>• Low latency<br>• High scalability | • Low reliability<br>• High complexity |
| | [41] | Enabling VFC in smart cities for traffic management | R | N | Not mentioned | • Low response time<br>• Low cost<br>• Low network traffic<br>• High scalability | • Low reliability<br>• Hard mobility management<br>• Low computational capability |
| Energy | [42] | Energy management using meta-heuristic algorithms in fog-based smart cities | S | M | Not mentioned | • Low energy<br>• Low latency | • High complexity<br>• Low reliability |
| | [43] | Fog-based smart cities in cooperation with mobile edge networks | S | N | Not mentioned | • Low complexity<br>• Low energy<br>• Fast convergence | • Low availability |
| Monitoring | [44] | Proposing a distributed framework for surveillance videos in fog-based smart cities | R | N | Apache Spark, Hadoop | • High scalability<br>• Low cost<br>• Low energy<br>• Low bandwidth | • Low availability<br>• Low security<br>• Low reliability |
| | [45] | A fog-cloud-edge based architecture for smart city applications | D | - | Java application on Tommcat, Docker, FIWARE, Java OpenJDK, JMeter, Ubuntu | • High scalability | • Vulnerable security |
| Security | [46] | A fog and cloud-based smart city middleware | P | - | JAVA platform | • High security<br>• High flexibility | • Low extensibility |

Binary Particle Swarm Optimization (BPSO) have been investigated to balance the requests and energy on virtual machines (VMs) of fog and environment. Although this paper evaluated response time, computational cost, and processing time via applying meta-heuristic algorithms, it suffers from high complexity. About mobile networks cooperating with fog computing, Dong, et al. [43] designed a system that cooperated with fog computing in order for the offloaded workload to be processed by forwarding data on the whole fog layer. A standard optimization issue of energy and QoE was devised in fog computing process integration with fairness.

About video summarization with the aim of monitoring, Nasir, et al. [44] represented a novel fog-based framework to summarize videos in a distributed way with the aim of video surveillance and monitoring in a smart city. The disadvantage of this paper was that no summarization algorithm was explored for limited-resource devices. Souza, et al. [45] proposed an architecture for categorizing and monitoring the applications of micro-services and services, which is implemented in a smart city environment. The proposed hybrid architecture brought horizontal scalability, but it suffered from security weaknesses.

As for security, Mohamed, et al. [46] presented SmartCity-Ware SOM (service-oriented middleware) for combining and using fog and CoT. To improve integration and to use different services; The SOM has abstracted services and features involved in applications of smart cities' assets of services that were available by the service-oriented pattern. The main benefit of this procedure was that the flexibility in expanding middleware was to provide new services for software at the time of development.

## 5.2 Resource-Based Approaches

This section represents resource-based class, including 9 reviewed papers characterizing approaches such as scheduling [47], [48], [49], resource provisioning [50], [51], resource allocation [52], [53], load balancing [54], and offloading [55]. For every paper in this category, we have represented information as a group, evaluation method, algorithms type, tool, advantages, and disadvantages in Table 6 for a better comparison. The following section shows the reviewed papers in resource-based approaches.

### 5.2.1 Review of Resource-Based Approaches

For coupling resource management, Wang, et al. [47] presented a model to resolve scheduling issues. This issue occurred when a sensor received numerous commands for servicing at the same instance, and it led to collisions and failures. This model was fog-based, and it extended the Hungarian method for management, which could cause delays in recognizing effective services. Liu, et al. [48] proposed ADGTS (adaptive double fitness genetic task scheduling) algorithm by collaborating task scheduling with fog resources in smart cities. ADGTS optimized the cost of communication and makespan of tasks at the same time. However, the authors did not investigate metrics such as availability and scalability. Concerning the interoperability and scheduling goals in smart cities applications, Rampérez, et al. [49] represented a fog-based architecture for smart cities to bring flexibility and decrease latency. Indeed, this architecture created various domains in the environments of a smart city like smart grid, smart traffic, etc. that exchanged information by a lightweight context broker who provided low-latency. The context broker depended on E-SilboPS, which is a flexible context-aware middleware.

Bartolí, et al. [50] expressed the advantages of fog and cloud coordination for a particular application on a smart city scenario, which needed a strategy for provisioning. The advantages did not just concentrate on particular service but also concentrated on the effect it had on a company that deployed services. The aim of this paper was not dealing with deploying architectures supporting management strategies but specifying the advantages that deployment of such architecture brought. Liao, et al. [51] established a fog-based vehicle mobility model according to geographical migration to compute resources in a smart city through satellites. In order to provide resources for vehicular computing based on the mobility migration pattern of geographical, the status of the roads was quantified accurately. Although, evaluation parameters such as scalability, security, and reliability of the proposed model were not assessed.

Toward resource allocation, Naranjo, et al. [52] proposed FOCAN (fog computing architecture network), a multi-layer architecture to provide an environment in which programs could route, communicate, and compute together in smart cities. The presented architecture reduced latency and service usage in comparison with other architectures with varied capabilities. Liao, et al. [53] presented a fog-based geographical movement model for resource allocation according to vehicle mobility in smart cities. In this model, the vehicles linked to the fog nodes to share the additional resources as resource computation suppliers. The vehicle as a service framework enjoyed the randomness of resource computation and enhanced the flexibility of centralized cloud architecture.

Relevant to balancing the workload in fog-based environments, Liang, et al. [54] presented a smart algorithm according to the Hungarian technique. In the first step, each fog node received the data from the nodes in the neighborhood; then, in order to find a solution for optimization, a genetic algorithm was designed. In the end, any fog node could decide to balance the workload of others with forwarding their workload. The result of the simulation process was that the mentioned algorithm decreased energy utilization and shortened delay. Alli and Alam [55] represented a secure computation offloading scheme in fog-Cloud-IoT (SecOFF-FCIoT) to reduce latency and energy usage to the smallest possible. The Neuro-Fuzzy scheme was used for resource validation, and besides, the particle swarm optimization (PSO) optimized the fog node selection process. Implementing results demonstrated that the delay in this model was low, and the energy usage insignificant. Hence, this result made the introduced model robust.

## 5.3 Application-Based Approaches

The application-based approaches are arranged to find and choose fog-based applications in a smart city. These techniques fall into different domains, namely Vehicle [56], [57], [58], noise detection system [59], [60], self-regulated system [61], hospitality industry [62], smart lighting [63], and smart building [64], to enhance the quality in smart cities. A comparison of every application-based approach is presented in Table 7, including group, evaluation method, algorithm type, tools, pros, and cons. In Section 5.3.1, we have reviewed the application-based approaches.

### 5.3.1 Review of Application-Based Approaches

Wang, et al. [56] presented a framework for IoV (internet of vehicle) based on information-centric and fog computing integration in a smart city to supply mobility by varied schema allocation taking data characteristic into account. Nevertheless, the framework was not feasible because the technology of fog not fully developed. Besides, the critical metrics were not evaluated like the scalability of this framework, security, and time. Caminha, et al. [57] presented an all-purpose system as SensingBus, which gathered all data generated by city buses. In the validation process, the authors built a prototype and examined that leastways 20 sensors could be accommodated by a fog node simultaneously. Expect these, no coordination among buses was performed to remove or decrease the duplicate information. On the specification of vehicle capability, Kui, et al. [58] determined the computation capability in VFC, such as the speed of the processor and the receiving tasks time. With this model, the characteristics such as time and place of the computing ability were revealed. Also, the connected vehicle distribution and the computing ability distribution were portrayed.

TABLE 6
A COMPARISON OF RESOURCE-BASED APPROACHES
In the evaluation method column, Prototype=> P, Simulation=>S, Real-testbed=>R, and Design=> D
In the algorithm type column, Non-Heuristic=>N, Metaheuristic=>M, and Heuristic=>H.

| Group | Ref | Main Focuses | Evaluation Methods | Algorithm Types | Tools | Advantages | Disadvantages |
|---|---|---|---|---|---|---|---|
| Scheduling | [47] | Coupling resource management in fog-based smart cities | S | N | Not mentioned | • Low coupling computing<br>• High resource utilization | • Low reliability<br>• Low security<br>• Low scalability |
| | [48] | A task scheduling algorithm in fog-based smart cities | S | M | Not mentioned | • High flexibility<br>• Low communication cost | • Vulnerable security<br>• High complexity |
| | [49] | Fog-based smart cities architecture with multi-domain standards | R | - | OpenStack (FI-WARE Lab) | • Low latency<br>• High scalability<br>• High throughput<br>• High interoperability<br>• High flexibility | • Low availability<br>• Low security |
| Provisioning | [50] | Resource management based on fog strategy in smart cities | R | - | Not mentioned | • High scalability<br>• High reliability<br>• Low latency | • Low availability |
| | [51] | Vehicle mobility-support migration for satellite-enabled fog-based smart cities | S | H | Not mentioned | • Real-time responding<br>• Low computation power | • low scalability |
| Allocation | [52] | Managing IoT applications in a fog-based architecture in smart cities | S | H | Not mentioned | • Low energy usage<br>• Low latency<br>• Effective management | • Low availability<br>• Low scalability |
| | [53] | Improving the efficiency and task processing via fog resource migration in smart cities | S | M | TOPSIS | • High flexibility<br>• Low energy<br>• Low response time | • Low security<br>• Low scalability<br>• High complexity |
| Load-balancing | [54] | Workload balancing in the cloud and fog-based smart cities via a distributed smart Hungarian algorithm | S | M/N | MATLAB | • Low energy<br>• Low latency<br>• Fast convergence | • Low heterogeneity<br>• High complexity |
| Offloading | [55] | Proposing a secure offloading scheme in fog and cloud-based smart cities using machine learning | S | - | NS-3, JAVA Platform | • High security<br>• Low latency<br>• Low energy<br>• High scalability | • Low availability |

Through providing a simulation model for the sound classification, Baucas and Spachos [59] investigated the productiveness of applying a balanced cloud and fog computing combination in urban sound classification system. So as to create a configuration that can make the sensing system more scalable, the authors concentrated on the trade-off and communication between server-side delay, application runtime, and power consumption of end-devices. In a similar work, Baucas and Spachos [60] presented a scalable framework based on IoT and fog computing for the urban sound system. The statistic results illustrated that this framework could enhance the scalability of the cloud-fog-based (hybrid architecture) urban sound sensing. However, the authors did not investigate the flexibility and cost of the framework.

By self-regulated systems in future smart cities, Abbas, et al. [61] presented a framework to develop intricate fog-based smart cities on the basis of emerging intelligent systems. The self-regulating system, in combination with fog computing, was introduced to advance applications of smart cities as a technological model in order to develop the scalability and complexity of future smart cities.

Kansakar, et al. [62] introduced an architecture based on fog computing for hospitality and related industry to make a good experience for guests, increment incomes, and improve business insights. The results verified 37% improvement of offloading tasks in fog-based architecture over cloud architecture. However, due to different peak hours for fog nodes, the architecture was not adaptive and interoperable enough to make better the efficiency of hospitality at peak hours. Concerning smart lighting, Jia, et al. [63] introduced a smart street lamp (SSL) architecture based on fog to conserve energy and safety of smart cities. The strength points of SSL architecture are listed in the following: 1) well management, independent operation on every single lamp of streets, 2) smart sensing, 3) real-time communications have made the network more efficient. The introduced architecture removed ultimately not required periodic checkups for human resources reduction.

TABLE 7
A COMPARISON OF APPLICATION-BASED APPROACHES

In the evaluation method column, Prototype=> P, Simulation=>S, Real-testbed=>R, and Design=> D
In the algorithm type column, Non-Heuristic=>N, Metaheuristic=>M, and Heuristic=>H.

| Group | Ref | Main Focuses | Evaluation Methods | Algorithm types | Tools | Advantages | Disadvantages |
|---|---|---|---|---|---|---|---|
| Vehicle | [56] | Fog-based mobility support for information-driven IoV in smart cities | S | - | Not mentioned | • Low delay<br>• Low drop rate | • High cost |
| | [57] | Utilizing bus lines in fog-based smart cities | P | - | OpenStack, Apache 2.4.18, MySQL 5.7, Django | • Low cost<br>• High scalability<br>• High security<br>• High availability | • Low reliability<br>• Low extensibility |
| | [58] | Specifying the abilities of VFC in large-scope cities | S | N | Not mentioned | • High scalability<br>• Low complexity<br>• Low latency | • Low availability<br>• low extensibility<br>• Security not evaluated |
| Noise | [59] | A scalable fog-cloud computing model for the urban sound system based on IoT | S | H | Python, Ubuntu (ver. 16.4) | • High Scalability<br>• Low latency<br>• Low power consumption | • High cost<br>• High complexity |
| | [60] | Large scale fog-IoT-based framework for urban sound classification | S | - | Python (pysound), Ubuntu (16.4) | • High Scalability<br>• Low latency<br>• Low power consumption | • Low flexibility<br>• High cost |
| Self-regulated | [61] | Smart systems for sustainable cities based on fog computing and self-regulated systems | S | - | JADE agent platform | • High scalability<br>• High flexibility<br>• High performance | • Low security<br>• Low reliability |
| Hospitality Industry | [62] | A fog-assisted paradigm to develop hospitality and other related industries in smart homes | R | - | Linux virtual machine, Amazon Elastic Compute Cloud (EC2) | • Low cost<br>• Local and global analyzing<br>• Energy-efficiency | • Medium security<br>• Limited interoperability |
| Smart lighting | [63] | Smart lamps in the streets for smart cities based on fog computing | S | N | Not mentioned | • Low energy<br>• High reliability<br>• Low latency<br>• High security | • Low scalability<br>• Low availability |
| smart building | [64] | An architecture based on IoT and fog/cloud computing in smart cities | P | - | MySQL MongoDB | • Low energy<br>• High security<br>• Low cost | • Low availability<br>• Low reliability |

Dutta and Roy [64] proposed an incorporated architecture, including IoT, cloud, and fog, to expand cost-effective and energy-aware buildings in smart cities. The presented prototype lets users handle all devices intelligently with real-time responses. However, the prototype was affordable and energy-efficient, but important concepts such as availability, extensibility, and scalability have not considered.

# 6 ANALYSIS OF RESULTS

According to the process presented in Section 4, we have described the result of the SLR process in this section. Also, based on the research question in Section 4.1, we have answered to the RQ3, RQ4, and RQ5.

## 6.1 Overview of the Selected Studies

For a better examination on fog-based smart cities and put into words the state of them, we have considered the following complementary questions:
- What are the active groups and research communities in fog-based smart cities?
- What are the fora that researchers and authors published their papers related to the fog-based smart cities?
- When did research about smart cities based on fog computing become active in research communities?

### 6.1.1 Active Research Groups or Communities

To answer the first complementary question, we needed to consider the distribution of active communities or groups. As indicated in Table 8, only communities that represented at least two studies, along with their primary research focus, are selected and listed. A remarkable number of researches are associated with service-based approaches that have published in the Shanghai Jiao Tong University in China, Dalian University of Technology in China, University of Rhode Island in the USA, Brown University in the USA, and Collage of Information Technology in UAE. Researchers in the College of Computer Science and Technology in China and the University of Electrical Science and Technology of China have focused on resource-based approaches. Also, in the School of Engineering, University of Guelph in Canada, researchers concentrated more on application-based.

### 6.1.2 Publication Fora

According to Table 9, two categories as journal and conference papers along with publication channel, publisher, abbreviation, and count of our reviewed studies are specified. Among these 32 studies, three of them are published in journals such as IEEE ACCESS (impact factor: 3.745), two of them in IEEE IoT-J (impact factor: 9.936), IEEE TII (impact factor: 9.112), and JPDC (impact factor: 2.296). Moreover, most of the papers in Table 9, published in service-based and application-based communities.

TABLE 8
ACTIVE GROUPS AND COMMUNITIES AND RESEARCH FOCUS

| Affiliation | Study ID | Research Focus |
|---|---|---|
| School of Electronic Information and Electrical Engineering, Shanghai Jiao Tong University, China | [35], [50], [53], [56] | Resource-Based, Application-Based, Service-Based |
| School of Engineering, University of Guelph, Guelph, Ontario, Canada | [59], [60] | Application-Based |
| University of Rhode Island USA | [4], [34] | Service-Based |
| Dalian University of Technology, Dalian, China. | [41], [62] | Service-Based, Application-Based |
| School of Computer Science and Information Engineering, Zhejiang Gongshang University, Hangzhou, China | [37], [47] | Service-Based, Resource-Based |
| College of Computer Science and Technology, Huaqiao University, Xiamen, China | [47], [54] | Resource-Based |
| University of Electrical Science and Technology of China | [34], [48] | Service-Based, Resource-Based |
| Brown University USA | [4], [34] | Service-Based |
| College of Information Technology, UAE University, Al Ain, UAE | [37], [46] | Service-Based |

TABLE 9
REVIEWED STUDIES DISTRIBUTION BASED ON PUBLICATION CHANNEL

| Category | Publisher | Publication Channel | Count |
|---|---|---|---|
| Journals | IEEE | IEEE Access | 3 |
| | | IEEE Internet of Things Journal (IEEE IoT-J) | 2 |
| | | IEEE Transactions on Industrial Informatics (IEEE TII) | 2 |
| | | IEEE Transactions on Sustainable Computing (T-SUSC) | 1 |
| | | IEEE Wireless Communications (IEEE-WCM) | 1 |
| | | IEEE Cloud Computing | 1 |
| | Science Direct | Journal of Parallel and Distributed Computing (JPDC) | 2 |
| | | Journal of Network and Computer Applications (JNCA) | 1 |
| | | Computer Communications (COMCOM) | 1 |
| | | Sustainable Computing: Informatics and Systems (SUSCOM) | 1 |
| | | Simulation Modelling Practice and Theory (SIMPAT) | 1 |
| | | Internet of Things (IOT) | 1 |
| | Springer | Peer-to-Peer Networking and Applications (PPNA) | 1 |
| | | Mobile Networks and Applications (MONE) | 1 |
| | Wiley | Wireless Communications and Mobile Computing (WCMC) | 1 |
| Conferences | IEEE | IEEE Global Communications Conference (GLOBECOM) | 1 |
| | | IEEE International Conference on Smart Energy Grid Engineering (SEGE) | 1 |
| | | IEEE International Smart Cities Conference (ISC2) | 1 |
| | | International Conference on Frontiers of Information Technology (FIT) | 1 |
| | | IEEE World Forum on Internet of Things (WF-IoT) | 1 |
| | | IEEE Symposium on Computers and Communications (ISCC) | 1 |
| | | IEEE International Conference on Communication Technology (ICCT) | 1 |
| | | International Wireless Communications & Mobile Computing Conference (IWCMC) | 1 |
| | | International Conference on Cloud Computing, Data Science & Engineering (Confluence) | 1 |
| | Springer | Parallel Processing Workshops (Euro-Par) | 1 |
| | | Software Engineering Methods in Intelligent Algorithms (CSOC) | 1 |
| | ACM | The ASE Big Data & Social Informatics (ASE BD&S) | 1 |

Fig. 4 illustrates the distribution of published papers in every single year along with their publishers. We observe that the earliest work was published in 2015, and the results remarkably raised in 2017. Based on the statistics, it is clarified that in the time range of 2013 to 2020, ACM has published one paper in 2015 and the very first paper in the field of fog-based smart cities. IEEE has released eight papers in 2017, five papers in 2018, and seven papers in 2019. ScienceDirect in the same time range has published one paper in 2018, four papers in 2019, and two papers in 2020. Also, Springer has published four papers, and Wiley published one paper in 2018.

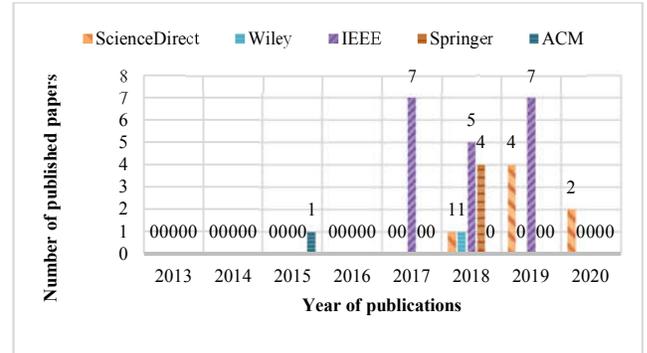

Fig. 4. Percentage of papers based on publishers

## 6.2 A Comparison and Classification of Fog-based Smart cities techniques

The results of our study are summarized to answer the research questions. Answering RQ3 first requires mentioning that the classification is analyzed and assessed in reliance on momentous metrics such as availability, reliability, security, scalability, response time, execution time, cost, and energy. The pros and cons of all classes were presented in detail, and it shows what each category suffers from more. Hence, we proposed a statistical analysis considering the fog computing approaches in smart cities. We presented two figures which depict the overall percentage of evaluated parameters and the rate of evaluated parameters per category. As shown in Fig. 5, the latency and energy were paid more attention than the other parameters by 26% and 18% in order. Response time by 12%, cost by 6%, and scalability improved by 16%. Security and throughput parameters have evaluated by 4% and 5% in order, which means they need to get paid attention more. Other parameters like reliability, availability, flexibility, and execution time had regarded only by 13%.

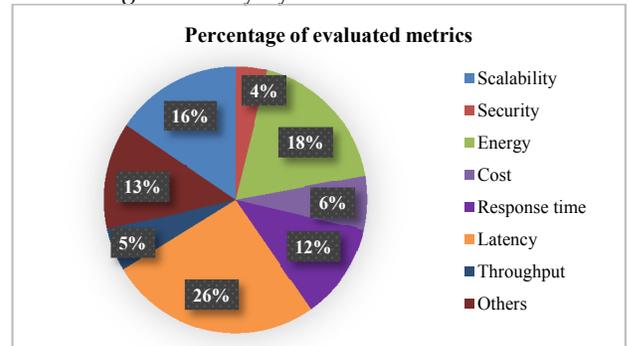

Fig. 5. Percentage of evaluated parameters in reviewed papers

In Fig. 6, we proposed the percentage of assessed metrics in every single class. In the resource-based approaches, the re-searchers reduced latency and energy by 26.1% equally, response time by 19.2%, and scalability by 8.7%. Nonetheless, only 4% of the authors have considered security in their work, and 8.7% of the authors assessed cost in. In the service-based approaches, all parameters were evaluated, and among them, 20% of the authors have tried to reduce latency, and 15% of them improve scalability. Other metrics such as reliability, availability, and execution time have evaluated by 15% in this class. In application-based approaches, the focus was on latency and scalability, in which 30% of the papers have covered latency, and 25% of them have assessed scalability. Nevertheless, security was left uninvestigated in this category.

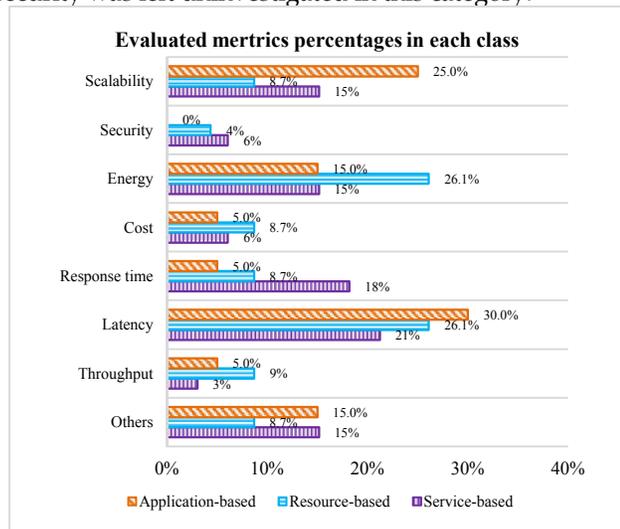

Fig. 6. Percentage of evaluated parameters per class

Regarding the tools that researchers have used in their works, as it is shown in Tables 5, 6, and 7, Top five of them are FIWARE, OpenStack, Java platform, Hadoop, and Python programming language. Also, the most popular databases that the authors used are MySQL 5.7 and MongoDB.

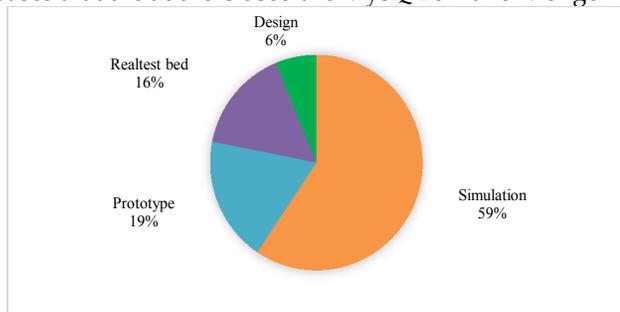

Fig. 7. Percentage of evaluation method of selected papers

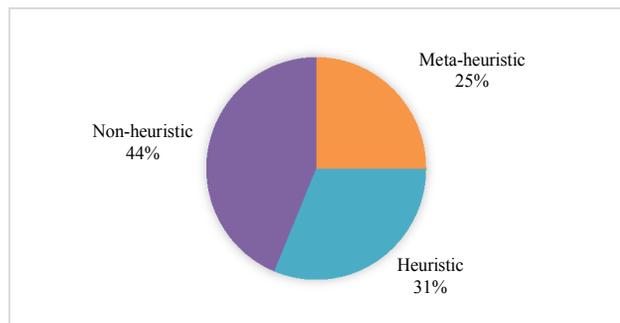

Fig. 8. Distribution of algorithm type

We answered RQ4 by demonstrating the evaluation method of all selected papers in Fig. 7. Researchers simulated 59% of their works, and only 16% of their work is tested in real. It can bring this issue to the table that more than half of the papers were simulated, and how can we say it will get the same result in a real environment? Besides, Fig. 7 shows that 19% of papers presented prototypes, and 6% of them were designed.

Answering RQ5, based on the statistics illustrated in Fig. 8, distribution of algorithm types includes 44% of non-heuristic algorithms, 25% of meta-heuristic algorithms, and 31% of heuristic algorithms.

# 7 OPEN ISSUES AND FUTURE TRENDS

We intend to discuss the challenges and issues which have not been perused in the fog-based smart cities in this section. As a response to RQ6: What are the open issues and future trends of fog-based smart cities? We have proposed the following answers

### 1) Optimal Solutions and Algorithms

The optimization algorithms used in smart cities based on fog computing approaches bring lower energy consumption, delay, and convergence for the improvement of cities. However, from the slant of complexity, these algorithms are regarded as a subclass of NP-hard or NP-complete class, such as scheduling [48] and geographic routing [65]. Consequently, there are other meta-heuristic or heuristic methods to reach the optimize solution such as simulated annealing [66], particle swarm optimization [67], gray wolf optimizer [68], bee colony [69, 70], bat algorithm [71], memetic algorithm [72-74], genetic algorithm [75], lion optimizer algorithm [76], and ant colony [75, 77]. All these solutions are engrossing in future studies.

### 2) Big Data Analytics

A massive collection of diverse and complex data that is being unprecedentedly stored and collected is called big data. Big data have caused problems and issues such as analysis, storage, and processing or result visualization. The five prominent attributes of big data are variety, volume, value, veracity, and velocity, which are known as 5Vs. [78]. Hence, any significant data movement from smart devices and equipment to the cloud may not be efficient. Plus, big data analysis is necessary for getting real-time and optimal responses in smart cities [34, 79]. They need to be processed in a short time with the lowest delay as possible. Therefore, getting real-time responses depends on finding the closest fog node to communicate with, then it comes to fog computing as an advantageous technology [80]. Big data analysis can be essential enough for the future of smart cities.

### 3) Green Smart Cities and Energy Consumption

Given the expansion of smart cities, the variety and scale of services provided in smart cities are one of the challenges facing energy resources [81]. Therefore, services provided in smart cities to reduce energy consumption should be optimized. Fog technology is one of the most appropriate tools to reduce energy consumption. Optimal energy consumption in smart cities is also one of the best ways to reduce greenhouse gas emissions. Therefore, research into creating "green" cities that have the least energy consumption [81] is one of the open issues ahead.

### 4) Social networks for management and surveillance

The inclusive of online social networks (OSN) in villages, cities, and countries has provided an excellent platform for their use in smart cities based on fog computing. Connecting people in cities through social networks has created a large graph. Also, the content transmitted through this vast network generates rich data. OSN can be used to predict how fog resources are distributed in smart cities. It also can predict on-demand services and how to apply for resources [82, 83]. Using the information and data processing created in OSN can create a different experience of smart cities based on social relationships [84]. Social networks such as Instagram, Twitter, and Facebook have created massive networks and data [85, 86]. Therefore, one of the best areas for future work is to use OSN in smart cities based on fog computing.

*5) Application Challenges*

The use cases and applications of fog-based smart cities are designed to be applicable in quite a few fields. In this subsection, we intend to look at a couple of applications that can be captivating for future works.

- *Smart transportation: traffic control and autonomous cars*

Coupling smart infrastructure with intelligent vehicles proves that a cloud-only scheme is not enough for autonomous transportation, and the fog computing model is needed. The Internet of Vehicles (IoV) emerged from IoT and mobile Internet convergence, where vehicles are intelligent moving nodes in a network [87, 88]. To specify the areas of transportation, we can mention boats or ships, trucks, trains, drones, and busses, and all of them require traffic control. The systems of controlling traffic receive information from other resources like an intelligent traffic light, cloud-based, and municipal systems. Accordingly, the data flow between the fog nodes, vehicles, and system of traffic control in any direction should certify that all equipment will have the control capability they require. Nevertheless, there are points left uninvestigated that can be interesting in future works, such as the effect of vehicle velocity on fog nodes' computational functionality. Plus, researchers did not check if there is any issues or challenges in large cities all over the world.

- *Smart homes*

The technology of smart homes is now a trend in current societies that supplies smart environments for living, along with comfort and convenience [89, 90]. Smart homes might have numerous sensors or devices working together, including occupancy, humidity, security, temperature, keycard readers, air quality, and door close/open. They are multitudinous issues for the future of smart homes, which enable them to be more efficient. The issues include energy consumption and carbon footprints, privacy and security, latency, failure management, and fault tolerance. They all can be interesting open trends.

- *Smart health-care*

One of the most engrossing areas in fog computing based smart cities is health-care and patient care. The critical information about people's health and medical status must transfer without latency and be secure and independent. At the same time, high-quality care requires to be maintained while the reduction in cost is demanded [10, 91]. Additionally, other issues to research are energy utilization resulting from task offloading, data management, scalability, cost, and interoperability [92]. They all are still open issues and future trends for researchers.

- *Smart surveillance systems*

Surveillance systems are an integral part of smart cities. Mainly surveillance is in areas where there is enough light with video recording devices or through the night-vision cameras that are controlled by human resources [93]. One of the significant challenges in smart cities is the integration of surveillance systems. Therefore, due to the vastness of municipal services as well as the size of the city, it is necessary to provide low latency and real-time systems by providing computing resources like fog nodes in cases such as emergencies [6]. Also, intelligent surveillance systems in smart cities based on fog can by themselves be one of the researches to do in the future.

- *Smart parking or shopping*

One of the interesting open issues that have plagued smart cities is the problem with the parking. There is usually a shortage of parking in urban areas, but predictable events such as events and festivals can add to the problem [94]. A large amount of data transfer from different locations in the city creates an overload in the network. Also, shops can be the reason for the lack of parking. Therefore, even with the intelligent shopping store, the location can partially solve the problem of lack of parking space. One of the fascinating future works is smart parking in fog-based smart cities.

*6) Security and privacy*

Privacy and security are a big challenge in smart cities. Smart sensors and computing nodes should have the security necessary to prevent any potential misuse, especially in critical environments. Also, privacy is another challenge for smart cities. Protecting the privacy priorities of service providers and users is essential to maintaining it. Due to the integration of various services in smart cities, providing services to users or providers can be associated with mistakes, such as sending information to people who are not authorized [95, 96]. However, according to our survey, a few papers have adequately tested and evaluated this critical metric. Therefore, one of the significant challenges in fog based smart cities is assessing security and privacy.

*7) Mobility challenges*

In the case of mobility, many fog nodes and IoT devices could be mobiles, tablets, vehicles, etc. the ubiquity growth of sensor devices is causing a spacious domain of delay-sensitive and large-scale applications [7]. Therefore, if the future location of devices and fog nodes can be predicted in a smart city, it helps to reduce total latency in the fog network and data forwarding. The result of mobility can be seen in fog-based smart cities by recording the pattern of mobility and behavior. Enhancing solutions of smart cities based on fog computing, the semantics of data, and contextual information can be extracted. The context-aware and data semantics applications need the location of equipment and users, and the capability of them [97]. The location- and context-aware computation in fog-based smart cities is an engrossing area for future researches.

*8) Other issues and challenges*

♦ **QoS:** While investigating and analyzing papers, we observed that QoS factors were not concurrently considered in fog computing-based smart cities. For instance, in some techniques such as scheduling and provisioning, time and cost, are evaluated, while other factors, namely reliability, and energy, are ignored. In order to make a trade-off between QoS factors, presenting a productive multi-objective technique can be a promising solution. Therefore, this area is an appealing research area for future works.

◆ **Scalability:** In smart cities based on fog, services should be implemented and tested on large-scale networks. However, most of the papers we reviewed did not meet critical criteria, such as scalability in their studies. Therefore, it is necessary to test and implement these services and systems on the real scale of a smart city to check their performance correctly. Investigating scalability in smart cities as well as implementing services in a real environment is one of the challenges smart cities can deal with.

◆ **Real-testbed:** It obviously can be seen that simulations are the primary evaluation method regarded by researchers. According to our review, 59% of the reviewed papers gained their results by using simulation environments. In fog-based smart cities, presumably, so many variations and changes happen when something like a service, model, prototype, or architecture is deployed. They might need upgrades, or their policies may change. So, dealing with these variations desire adaptation while, in this context, the real-testbed results and adaptability were not paid enough attention and can be another challenge.

## 8 THREATS TO VALIDITY AND LIMITATIONS

The systematic literature reviews require to be accurate and complete as possible, and, in this work, we have tried our best to provide an SLR with minimum limitations. Although an SLR is giving reliable information, it has its potential restraints. In this section, we discuss them and other threats to our work in different stages of this SLR.

*Identification Threats to Selected Studies:* To avoid any bias, we attempted to retrieve the current literature as possible as they can during our search steps and strategies. Moreover, determining the scope of this study was another challenge due to the relevance of different communities or fields, including IoT and cloud computing, to fog-based smart cities. There are different phraseologies for the same concepts that communities use. Hence, we had to avoid any bias and cover all the aspects to search for standard terms and gather them in our search string. This approach prevented bias but increased search work. In this work, several rich databases, including IEEE, ACM, Springer, ScienceDirect, Wiley, and Google scholar are used as trustable sources. Although the use of these renowned electronic databases will lead to the most relevant and authoritative result, it cannot guarantee that selecting all proper studies has happened. Therefore, some papers were removed due to the mechanism described in Section 4.

*Selection Threads and data extraction:* Quality assessment is a necessity for every study in SLRs [16]. Therefore, based on the research methodology and paper selection process (Section 4 and Section 4.2), the most qualified and reliable studies selected for our first study. For the most authoritative and qualified papers, only journals and conference papers are included in our review. With all respect to the suitability and ability of non-reviewed papers, magazines, non-English papers, and books, they have not been considered. Additionally, we had to select those papers which had answers to our main research questions. A precise, systematic review requires quality evaluation to evaluate the impact of the primary studies' quality on our conclusions.

## 9 CONCLUSION

This paper systematically reviewed the existing fog-based approaches in smart cites. Initially, we had an overview of the status of fog computing in smart cities. We attempted to indicate the uninvestigated areas by putting gained information together and represent them as future research. Hence, to show the result, we proposed visual diagrams and tables by virtue of transferring information among practitioners and researchers in the field of fog-based smart cities. In this paper, by applying our search query and based on inclusion/exclusion criteria, 98 papers from 2013 to 2020 were selected. After that, according to the explained procedure in research methodology, 32 papers were selected for a more accurate review. We then classified the selected papers into three main classes: resource-based, service-based, and application-based for a closer look. Each of these classes was meticulously evaluated with metrics such as scalability, security, energy, cost, response time, latency, and throughput. It should be noted that almost 43% of papers were service-based, 28% resource-based, and 28% were application-based. Most concerns were about latency and energy consumption by 26% and 18% in order. Additionally, most papers did not analyze some important metrics properly, such as security and throughput. During the SLR, we found that 44% of the algorithm types were non-heuristic, 31% of them were heuristic, and 25% of algorithm types were meta-heuristic. Furthermore, on the way to the smart city, the ahead controversial challenges and future work for researchers to study were outlined.

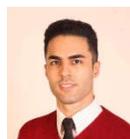
**Morteza Rahimi** is a researcher at Young Researchers and Elite Club. He received his B.S. in Computer Engineering from Shahr-e-Qods Branch, IAU, Tehran, Iran, in 2020. Morteza is also a research assistant and teaching assistant in the Department of Computer Engineering at the IAU University. He has previous experiences in peer-reviewing and publishing papers in technical journals. His research interests include fog computing, machine learning, big data, and the Internet of Things.

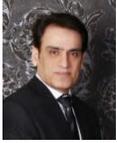 **Mostafa Haghi Kashani** received his BS in Computer Engineering from Kashan Branch of IAU, Iran, in 1999 and the MS in Computer Engineering from South Tehran Branch of IAU, Iran in 2002. He is currently a full-time PhD Candidate in Computer Engineering-Software Systems at Science and Research Branch of IAU, Tehran, Iran. He is a researcher and lecturer in the Department of Computer Engineering at the IAU University. He has authored/co-authored several papers in technical journals and conferences. His research interests include distributed systems, fog computing, social networks and evolutionary computing. He has acted as a reviewer in several international journals, including the Journal of Supercomputing (Springer), the International Journal of Communication Systems (Wiley), and the Journal of Big Data (Springer).